\documentclass[oneside]{article}
\usepackage{graphicx, natbib} 
\usepackage{url}
\usepackage{parskip}
\usepackage[a4paper, margin=25mm,centering]{geometry}

\title{The super learner for time-to-event outcomes: \\ A tutorial}
\author{Ruth H. Keogh$^{1*}$, Karla Diaz-Ordaz$^{2}$, Nan van Geloven$^{3}$, \\ Jon Michael Gran$^{4}$, Kamaryn T. Tanner$^{5}$}

\date{\small {$^{1}$ Medical Statistics Department and Centre for Data and Statistical Science for Health, London School of Hygiene \& Tropical Medicine, London, UK
\\
$^{2}$ Department of Statistical Science, University College London, London, UK
\\
$^{3}$ Department of Biomedical Data Sciences, Leiden University Medical Center, Leiden, NL
\\
$^{4}$ Oslo Centre for Biostatistics and Epidemiology, Department of Biostatistics, Institute of Basic Medical Sciences, University of Oslo, Oslo, Norway
\\
$^{5}$ Butler Columbia Aging Center, Mailman School of Public Health, Columbia University, United States
\\[2pt]
$^{*}${Corresponding author: ruth.keogh@lshtm.ac.uk}\\
}}

\begin{document}

\maketitle

\begin{abstract}
Estimating risks or survival probabilities conditional on individual characteristics based on censored time-to-event data is a commonly faced task. This may be for the purpose of developing a prediction model or may be part of a wider estimation procedure, such as in causal inference. A challenge is that it is impossible to know at the outset which of a set of candidate models will provide the best risk estimates. The super learner is a powerful approach for finding the best model or combination of models (`ensemble') among a pre-specified set of candidate models or `learners', which can include both `statistical' models (e.g. parametric, semi-parametric models) and `machine learning' models. Super learners for time-to-event outcomes have been developed, but the literature is technical and the full details of how these methods work and can be implemented in practice have not previously been presented in an accessible format. In this paper we provide a practical tutorial on super learner methods for time-to-event outcomes. An overview of the general steps involved in the super learner is given, followed by details of three specific implementations for time-to-event outcomes. These include the originally proposed super learner, which involves using a discrete time scale, and two more recently proposed versions of the super learner for continuous-time data. We compare the properties of the methods and provide information on how they can be implemented in R. The methods are illustrated using an open access data set and R code is provided.
\end{abstract}

\section{Introduction}
\label{sec:intro}

This paper focuses on methods for estimation of risks or survival probabilities conditional on individual characteristics based on time-to-event data. This may be for the purpose of developing a prediction model, in which the aim is to estimate the risk of an event occurring over a given time period conditional on a set of covariates. Estimation of conditional risks or survival probabilities can also be a step in a wider estimation procedure, where prediction is not the end goal. For example, in some methods for estimating causal effects of treatments, a step in the estimation procedure is to obtain survival probabilities conditional on covariates (e.g. \citet{moore_rcts_2011,westling_inference_2023,munch_statelearner_2024}). We focus on the setting in which the aim is risk prediction, but the methods described can also be used in wider estimation procedures of the types mentioned above. 

A challenge in prediction modelling, and more generally in estimating any conditional risk, is that it is impossible to know at the outset which of a set of candidate models will provide the best fit, this typically being measured in terms of the out-of-sample predictive performance. Analysts are faced with the task of choosing from a suite of different time-to-event models. Time-to-event data are very often subject to right censoring of event times, and analysis methods need to accommodate this \citep{andersen_analysis_2021}. The semi-parametric Cox proportional hazards model is widely used to develop risk prediction models \citep{cox_1972}. Parametric models such as the Weibull model are also used. Extensions include using the Cox model in combination with the Lasso \citep{tibshirani_lasso_1997}, ridge and elastic net regression approaches and generalised additive models (GAMs) \citep{bender_generalized_2018}. Non-parametric and machine learning approaches for risk prediction that avoid parametric (or semi-parametric) modelling assumptions have been developed, for example random survival forests \citep{ishwaran_random_2008}. Different methods for developing a prediction model are often applied, with the resulting models then being compared in terms of their predictive performance, but this can be ad hoc. The super learner (SL) is a powerful approach for finding the best model or combination of models (`ensemble') among a pre-specified set of candidate models or \emph{learners}, often referred to as the SL \emph{library} \citep{polleyrose_superlearner_2011}. The \emph{non-ensemble} SL selects the single best model and the \emph{ensemble} (or standard) SL finds an optimally weighted combination of the set of candidate learners, which is guaranteed to perform as well as the best performing learner included in the SL library. The \emph{non-ensemble} SL is often referred to as the \emph{discrete} SL, but we do not use this terminology in the current paper in order to avoid any confusion with a discrete-time implementation of the SL. The ensemble SL is a type of \emph{stacking algorithm}, which have a longer history \citep{wolpert_stacked_1992, breiman_stacked_1996}. We focus on the SL because of its flexibility to include parametric, semi-parametric and non-parametric methods, including machine learning methods of different `architectures', and because it has been shown to have the \emph{oracle property}. The oracle property says that the SL will perform at least as well as the best performing individual learner among the candidates, with this being true for both the ensemble and non-ensemble SLs. Unlike more ad hoc processes for identifying the best performing model from a set of candidates, which are often not pre-specified, the SL can be pre-specified as an approach to deriving a best-performing model. The derivation of a SL is based on minimizing a so-called \emph{loss function}, which is a function describing the difference between estimates from the model and true (observed) values.

The SL was originally proposed for continuous or binary outcomes, for which there are many models that could be included as candidate learners in the SL \citep{vdl_super_2007,polleyrose_superlearner_2011}. The SL was extended for use with time-to-event outcomes by \citet{polley_superlearner_2011}, who described a discrete-time approach. A discrete-time approach enables the use of candidate learners that are suitable for binary outcomes, but has the disadvantage that flexible estimation of the baseline hazard is challenging. More recently, two versions of the SL that accommodate continuous time have been proposed by \citet{westling_inference_2023} and \citet{munch_statelearner_2024,munch_jossl_2025}. By working in continuous time, these SLs can use candidate learners including Cox regression and its extensions, parametric survival models, random survival forests, and others. The different versions of the SL also differ in terms of how they handle censoring and in terms of the loss function used. SL methods for survival outcomes \citep{polley_superlearner_2011, westling_inference_2023,munch_statelearner_2024, munch_jossl_2025} have been described in a technical way and some of them are reported within papers whose primary focus is on using the SL within a wider causal inference estimation procedure. An existing tutorial focused on the SL for continuous or binary outcomes is available \citep{naimi_stacked_2018}, but does not extend to time-to-event outcomes. The aim of this paper is to provide an accessible tutorial outlining how to use the SL for prediction of time-to-event outcomes.

The paper is organised as follows. In Section \ref{sec:setup} we define the set-up and notation and give an overview of the general steps of the SL. In Section \ref{sec:discrete} we outline the SL that uses a discretised time scale, as described by \citet{polley_superlearner_2011}. Section \ref{sec:continuous} outlines two continuous-time versions of the SL, which do not require a discretisation of the time scale \citep{westling_inference_2023, munch_jossl_2025}. The practical implementation of the methods in R is discussed in Section \ref{sec:implementation}, and in Section \ref{sec:application} we illustrate the use of the methods using the `Rotterdam' data set, which is an openly accessible data set available in the survival package in R \citep{therneau_survival_2024}. We conclude with a discussion in Section \ref{sec:discussion}. To implement the methods we make use of R packages and Github repositories made available by the methods developers. We also provide simplified code that aims to provide a simpler illustration of the practical application of some of the methods. Code for reproducing the results in Section \ref{sec:application} is provided at \url{https://github.com/ruthkeogh/superlearner_survival_tutorial}.

\section{Set-up and overview of the super learner}
\label{sec:setup}

\subsection{Notation and set-up}
\label{sec:notation}

The focus is on a situation in which we have a data set $D$ of $n$ individuals available for the development of a prediction model. Let $T$ denote the time of the event of interest and $C$ the censoring time. The observed time is $\tilde T=\min(T,C)$, and $\Delta=I(T\leq C)$ is the event indicator. The set of covariates (predictors) that are candidates for inclusion is denoted $X$. The observed data for individual $i$ is $\{\tilde T_i, \Delta_i, X_i\}$. We assume a continuous-time setting, though in one of the SL methods considered time is discretised for the analysis. 

We let $S(t|X=x)=\Pr(T>t|X=x)$ denote the conditional survival probability for the event by time $t$. We will also need to refer to the conditional survival probability for censoring, which we denote $G(t|X=x)=\Pr(C>t|X=x)$. Our focus is on a situation without competing events. The target estimand is the conditional survival probability by a time horizon $\tau$, $S(\tau|X=x)$ (or 1 minus this, which is the risk or cumulative incidence). In the prediction context we are typically interested in being able to obtain an accurate estimate of $S(\tau|X=x)$ for a new individual that is not part of the training data $D$, and in the example application in Section \ref{sec:application} we will also consider a test data set.

\subsection{Overview of super learner steps}
\label{sec:overall.SL.steps}

Before outlining the details of specific implementations of the SL, we provide an overview of the general steps involved. We suppose that there are $p$ candidate learners for estimating the target estimand. We do not give details of different candidate learners, but as noted earlier these could include a mixture of parametric, semi-parametric and machine learning methods. The set of candidate learners can include methods of the same type but with different hyper-parameters, for example random forests with different maximum tree depth. It can also include methods that may perform some variable selection, such as the Lasso.

The SL involves use of cross-validation to fit and assess the performance of different models, and in the ensemble version of the SL to derive the ensemble. The general steps involved in fitting a SL are as follows, with further details provided below:
\begin{enumerate}
    \item {\bf Dividing data into folds}. Divide the data $D$ into $K$ folds, $k=1,\ldots,K$. Let $D_{-k}$ denote the $k$th training set ($k=1,\ldots,K$), which is the full data excluding the $k$th fold. Let $D_{k}$ denote the $k$th validation set ($k=1,\ldots,K$), which is the $k$th fold.
    \item {\bf Cross-validation  of candidate learners}. Fit each of the $p$ candidate learners to the training set $D_{-k}$, for $k=1,\ldots,K$. 
    \item {\bf Obtaining cross-validated estimates}. Use candidate learner $j$ fitted using $D_{-k}$ to obtain predictions of the target estimand in validation set $D_{k}$, for $j=1,\ldots,p$ and $k=1,\ldots,K$. This gives cross-validated estimates of the survival probability $S(\tau|X=x)$ for each individual in $D$. 
    \item {\bf Estimating the mean expected loss}. A \emph{loss function} is a measure of the difference between predictions of the target estimand and the true value. The value of the loss function is calculated for each individual using each candidate learner $j=1,\ldots,p$ in each validation set $k=1,\ldots,K$. The expected loss in validation data set $D_{k}$ using candidate learner $j$ (based on candidate learner $j$ fitted using $D_{-k}$ in step 3) is denoted $L^{(j,k)}$. The \emph{mean expected loss} across the $K$ validation data sets is calculated for each candidate learner, denoted $\bar L^{(j)}$, and this is sometimes referred to as the \emph{cross-validated risk} for candidate learner $j$. 
    \item {\bf Deriving an ensemble}. The \emph{non-ensemble} SL stops following step 4 and selects the candidate learner with the smallest cross-validated risk. The \emph{ensemble SL} uses further steps to derive an optimal combination or \emph{ensemble} of the candidate learners. Broadly, this involves fitting a regression model with the observed outcome as the dependent variable and with the predictions from each candidate learner (obtained in step 3) as the explanatory variables, e.g. if there are $p=10$ candidate learners then there will be 10 explanatory variables. 
\end{enumerate}

 The cross-validation used in the SL is crucial to guarantee the oracle property, as well as to avoid over-optimism. In step 2, the continuous-time methods that we will consider \citep{westling_inference_2023,munch_jossl_2025} also involve fitting models for the censoring distribution to each training set $D_{-k}$.  For step 4, one example of a loss function is the squared difference between the observed outcome and estimated survival probability, weighted to account for censoring \citep{polley_superlearner_2011}. Other loss functions will be discussed in the context of the different versions of the SL described below. See \citet{sonabend_examining_2024} for general discussion on loss functions for time-to-event outcomes. The cross-validated risk referred to in step 4 should not be confused with the risk of the event occurring, and we will use the alternative term \emph{mean expected loss} to avoid confusion. In step 5 the regression model used to derive the ensemble should be done on a scale that is compatible with the loss function and the regression coefficients need to be constrained to be non-negative and to sum to 1. One example is to use non-negative least squares, resulting in non-negative coefficients that can be thought of as weights given to each candidate learner, with a weight of 0 meaning that the corresponding candidate learner does not contribute to the SL ensemble. 

 For the ensemble SL the process results in the coefficients (or weights) for the candidate learners, whereas the non-ensemble SL results in the best single learner. To then use the resulting SL to obtain an estimate of the target estimand $S(\tau|X=x)$ for a new individual we need to refit the necessary candidate learners on the full data $D$. For the non-ensemble SL we only need to refit the selected learner, and for the ensemble SL we refit each of the candidate learners that is assigned a non-zero weight in step 5. The final ensemble SL combines estimates from these learners fitted on the full data with the coefficients estimated in step 5. 

The non-ensemble SL is essentially a ``standard'' prediction model chosen via cross-validation. However, when following the SL algorithm, this prediction model is obtained through a principled and pre-specifiable process of choosing among many different candidate learners, which can be of different types and architectures, as well as including learners of the same type but with different tuning parameter choices.



\section{Discrete-time super learner}
\label{sec:discrete}

\subsection{Discrete data set-up and notation}
\label{sec:discrete.setup}

\citet{polley_superlearner_2011} described a SL for survival outcomes, which involves using a discretised time scale. Time is discretised by dividing each individual's follow-up into a series of discrete time periods up to and including the period in which they have the event or are censored, with each period having an associated indicator of whether the individual had the event of interest in that period or remained event-free at the end of the period. This turns the problem of predicting a time-to-event outcome into that of predicting a binary outcome (the discrete-time event indicator) across a series of time points, using individuals still at risk at those time points, enabling use of any method suitable for binary outcomes in the set of candidate learners. The learners are then models for the discrete-time hazard. To simplify the notation of the methods described below we assume, without loss of generality, that follow-up time is discretised into periods of unit length, $[t,t+1)$ for $t=0,\ldots,\tau-1$. We let $\Delta(t)$ denote an individual's status at the end of period $t$, with $\Delta(t)=0$ for periods in which the individual remains at risk at the end of the period (or if the period ends with censoring) and $\Delta(t)=1$ for the period in which the individual has the event. This results in a data set in counting process format with several rows per person (in other words a `long' data set) - one row for each time period in which they are at risk. We denote the conditional discrete-time hazard at time $t$ by $Q(t|X)=\Pr(t\leq T<t+1|X,T\geq t)$. The conditional survival probability at time $\tau$ can be expressed as $S(\tau|X=x)=\prod_{t=0}^{\tau-1}1-Q(t|X=x)$. The conditional survival probability of interest can therefore be estimated based on estimates of the conditional discrete-time hazards. If $\tau$ is not one of the times in the discrete time grid (i.e. it is in between times on the grid) then $\tau$ is  replaced by $\lfloor \tau \rfloor$ in the product. 

\subsection{SL steps 1-3: Cross-validation}
\label{sec:discrete.cv}

For step 1 of the general SL algorithm outlined in Section \ref{sec:overall.SL.steps} the data are divided into $K$ folds. This should be done on the basis of individuals rather than on the basis of rows in the long data set, so that all the rows of data for a given individual are in the same fold. For step 2 we need to fit each candidate learner for the conditional hazard $Q(t|X)$ as a function of $t$. The outcome in these learners is $\Delta(t)$ (for $t=0,\ldots,\tau-1$) and the predictors are $t$ and $X$. The learners are fitted on the long data set, i.e. pooled across times. Candidate learners for $Q(t|X)$ could be any method for binary outcomes. In their illustrative application \citet{polley_superlearner_2011} included logistic regression models, random forests, generalised additive models (GAMs), polyclass, and a neural network. As noted earlier, the same type of learner can be used with different hyper-parameters, and \citet{polley_superlearner_2011} used GAMs with different degrees of freedom and logistic regression models with and without interaction terms. 

Fitting a model for $Q(t|X)$ requires consideration of how to incorporate time into the model, i.e. how to model the baseline discrete-time hazard as a function of time. Including separate indicators for each time period $t=0,\ldots,\tau-1$ could often be infeasible due to many time periods and/or too few events within some periods. Time could be included as a continuous variable in the candidate learners, with different functions of time used as separate variables. As an alternative, \citet{polley_superlearner_2011} proposed a two-stage procedure which involves first fitting a logistic regression including covariates $t$ (as a continuous variable) and $X$, and then using the fitted values from this as the offset term in a GAM including a flexible form for $t$. The degrees of freedom in the GAM used in the second stage is then a tuning parameter. The first stage model could alternatively be a random forest or a GAM, before using the resulting fitted values as the offset in a GAM including $t$ in the second stage. 

After fitting candidate learner $j$ in each of the $K$ training sets $D_{-k}$ ($j=1,\ldots,p$; $k=1,\ldots,K$), the resulting learner is used to obtain an estimate of the conditional hazard $\hat Q_j(t|X=x_i)$ for each individual $i$ in the validation set $D_k$ ($k=1,\ldots,K$). The cross-validated estimate of the survival probability based on candidate learner $j$ is obtained as $\hat S_j(\tau|X=x_i)=\prod_{t=1}^{\tau}1-\hat Q_j(t|X=x_i)$ for each person $i$ in $D$.

\subsection{SL step 4: Estimating the mean expected loss}
\label{sec:discrete.loss}

After fitting the candidate learners, the next step is to estimate their predictive performance based on a chosen loss function. As noted in Section \ref{sec:overall.SL.steps}, a loss function is a measure of the difference between estimates of the target estimand and the true value. The loss function is defined at a individual level. When the focus is on estimating the conditional survival probability it is natural to consider the squared difference between $S(\tau|X=x)$ and the event indicator $I(T >\tau)$ as the loss function: i.e. $\left(I(T >\tau )-S(\tau|X=x)\right)^2$. However, when there is right censoring we do not observe $I(T >\tau)$ for everyone in the data, and instead we observe $I(\tilde T >\tau)$ and $\Delta$. To account for censoring, a loss function that uses an inverse probability of censoring weight (IPCW) is defined as
\begin{equation}
    L_{IPCW}=\frac{\Delta I(\tilde T \leq \tau)}{G(\tilde T|X=x)}\left(0-S(\tau|X=x)\right)^2+\frac{I(\tilde T > \tau)}{G(\tau|X=x)}\left(1-S(\tau|X=x)\right)^2.
    \label{eq:disc.loss.IPCW}
\end{equation}
This is equivalent to individual contributions to the Brier score \citep{graf_assessment_1999}. Using this loss function requires an estimate of the conditional survival probability for censoring $G(\cdot|X)$, and we discuss this further below. \citet{polley_superlearner_2011} also considered alternative loss functions, which have the advantage of avoiding having to estimate the conditional censoring distribution, which is a nuisance parameter in (\ref{eq:disc.loss.IPCW}). The alternative loss functions are based on the hazard rather than the survival probability. One is a squared error loss summarised across all times:
\begin{equation}
    L_{L2}=\sum_{t}I(\tilde T\geq t)\left(\Delta(t)-Q(t|X=x)\right)^2,
    \label{eq:disc.loss.L2}
\end{equation}
and the other is based on the log likelihood:
\begin{equation}
    L_{loglik}=\sum_{t}I(\tilde T\geq t)\log\left( Q(t|X=x)\right)^{\Delta(t)}\log\left(1-Q(t|X=x)\right)^{1-\Delta(t)}.
    \label{eq:disc.loss.loglik}
\end{equation}
\citet{polley_superlearner_2011} comment that all three of the above loss functions are valid, but that (\ref{eq:disc.loss.IPCW}) is the most `natural' to use as it directly targets the target estimand, which is the conditional survival probability.

For each candidate learner, we estimate the expected loss in each of the $K$ data folds based on the chosen loss function. To estimate the expected loss based on $L_{IPCW}$ in (\ref{eq:disc.loss.IPCW}) requires an estimate of $G(\tilde T|X)$ or $G(\tau|X)$ (for individuals who remain under observation at time $\tau$). If the censoring is assumed to be independent then $G(t|X)$ can be replaced by the marginal probability $G(t)$, which can be estimated non-parametrically using the Kaplan-Meier estimator, and this is the approach taken by \citet{polley_superlearner_2011}. The censoring weights could alternatively be allowed to depend on $X$, for example through use of a SL to estimate $G(t|X)$. This would be performed separately from the main SL. The censoring probabilities are estimated in the validation data, i.e. using $D_k$. Assuming independent censoring, the expected loss for candidate learner $j$ in validation set $D_k$ is then estimated using 
\begin{equation}
    \hat L_{IPCW}^{(j,k)}=\frac{1}{|D_k|}\sum_{i\in D_k}\frac{\Delta_i I(\tilde T_i \leq \tau)}{\hat G(\tilde T_i)}\left(0-\hat S_j(\tau|X=x_i)\right)^2+\frac{I(\tilde T_i > \tau)}{\hat G(\tau)}\left(1-\hat S(\tau|X=x_i)\right)^2
    \label{eq:disc.loss.IPCWest}
\end{equation}

The expected loss based on $L_{L2}$ can be estimated using
\begin{equation}
    \hat{L}_{L2}^{(j,k)}=\sum_{t=0}^{\tau-1}\frac{1}{|D_k|}\sum_{i\in D_k}I(\tilde T_i \geq t)\left(\Delta_j-\hat{Q}_j(t|X=x_i)\right)^2,
    \label{eq:disc.loss.L2est}
\end{equation}
and the expected loss based on (\ref{eq:disc.loss.loglik}) is estimated similarly. The estimated mean expected loss for each candidate learner $j$ is then obtained by taking the mean across the $K$ folds: $\bar L^{(j)}=\frac{1}{K}\sum_{k=1}^{K}\hat L^{(j,k)}$, $j=1,\ldots,p$. 

\subsection{SL step 5: Deriving an ensemble}
\label{sec:discrete.ensemble}

The non-ensemble SL chooses the candidate learner with the best (i.e. lowest) cross-validated mean expected loss, based on the chosen loss function. In this section we focus on the ensemble SL, which derives an optimally weighted combination of the candidate learners, in such a way that the combination minimizes the chosen loss function. To give the general idea, first consider the simplified setting of a continuous outcome $Y$ and suppose we had fitted $p$ candidate learners resulting in cross-validated fitted values (conditional expectations) $\hat Y_{1},\ldots,\hat Y_{p}$ for each individual in the data. The SL ensemble would then be derived based on the model $E(Y|\hat Y_{1},\ldots,\hat Y_{p})=\alpha_1 \hat Y_{1}+\cdots +\alpha_p \hat Y_{p}$, with the parameters $\alpha_j$ ($j=1,\ldots,p$) estimated subject to the constraints that $\alpha_j\geq 0$ for all $j=1,\ldots,p$, and $\sum_{j=1}^{p}\alpha_j=1$. This can be achieved by fitting the regression model using non-negative least squares and then scaling the coefficients to sum to 1. This minimises the mean squared error, which is the appropriate loss function for a continuous outcome. 

For the time-to-event setting different choices for the loss function were summarised above, and these correspond to slightly different regressions to derive the SL ensemble. We begin by considering a SL based on the loss function $L_{L2}$ in (\ref{eq:disc.loss.L2}), as this is most similar to the simpler case of a continuous outcome outlined above. The aim is to obtain an algorithm from which we can obtain estimates of the conditional hazard $Q(t|X)$ in such a way that $L_{L2}$ is minimized. This can be achieved by a regression of $\Delta_i(t)$ on the estimates of the conditional hazard from the $p$ candidate learners, $\psi_{i,j}(t)=\hat Q_j(t|X=x_i)$ for $j=1,\ldots,p$. The model is fitted across all times $t=0,\ldots,\tau-1$ combined using the long data set. The regression equation is 
\begin{equation}
    E\{\Delta(t)|\psi_{1}(t),\ldots,\psi_p(t)\}=\sum_{j=1}^{p}\alpha_j \psi_{j}(t)
    \label{eq:SL.ensemble.L2}
\end{equation}
where the $\alpha_j$ ($j=1,\ldots,p$) are the regression coefficients (or weights) to be estimated. The coefficients $\alpha_j$ are constrained to be non-negative and to sum to 1, i.e. $\alpha_j\geq 0$ for all $j=1,\ldots,p$ and $\sum_{j=1}^{p}\alpha_j=1$. As in the continuous case, this can be achieved by fitting the model in (\ref{eq:SL.ensemble.L2}) using non-negative least squares regression, without an intercept. The resulting regression coefficients $\hat \alpha^*_j$ ($j=1,\ldots,p$) can then be scaled so that they sum to 1 using $\hat \alpha_j=\hat \alpha^*_j/\sum_{j=1}^{p}\hat \alpha^*_j$. The resulting SL estimator for the conditional hazard is 
\begin{equation}
     Q^{SL}(t|X=x_i)=\sum_{j=1}^{p}\hat \alpha_j  \psi_{i,j}(t),
    \label{eq:SL.est.L2}
\end{equation}
and these are then used to calculate the conditional survival probabilities of interest. 

If the chosen loss function is $L_{loglik}$ (equation (\ref{eq:disc.loss.loglik})) then a modified version of the above procedure is needed to derive a SL such that $L_{loglik}$ is minimized. \citet{polley_superlearner_2011} recommend transforming the conditional hazards using $g(\psi_{j}(t))=\mathrm{logit}\left \{\psi_{j}(t)\right \}$, and using the logistic regression equation 
\begin{equation}
    \mathrm{logit} \Pr(\Delta (t)=1|\psi_{1},\ldots,\psi_p)=\sum_{j=1}^{p}\alpha_j g(\psi_{j}(t)).
    \label{eq:SL.ensemble.loglik}
\end{equation}
They note that a constraint on the coefficients $\alpha_j$ ($j=1,\ldots,p$) is not required under this approach because the transformed conditional hazards $g(\psi_{j})$ are not constrained. The SL estimator for the conditional hazard based on $L_{loglik}$ is then 
\begin{equation}
    Q^{SL}(t|X=x_i)=\mathrm{expit}\left( \sum_{j=1}^{p}\hat \alpha_j g( \psi_{i,j}(t))\right).
    \label{eq:SL.est.loglik}
\end{equation}

Lastly we consider the case where the chosen loss function is $L_{IPCW}$ (equation (\ref{eq:disc.loss.IPCW})), which is based on the conditional survival probability rather than the conditional hazard. Let $ \phi_{j}=\hat S_j(\tau|X=x)$ for $j=1,\ldots,p$. The SL to minimize this loss function is derived based on the regression equation
\begin{equation}
    E\{I(T>\tau)|\phi_{1},\ldots,\phi_p\}=\sum_{j=1}^{p}\alpha_j \phi_{j}.
    \label{eq:SL.ensemble.IPCW}
\end{equation}
The regression can be performed using a weighted non-negative least squares regression of $I(\tilde T_i>\tau)$ on $\hat \phi_{i,1}, \ldots,\hat \phi_{i,p}$ among individuals who either: (1) have the event before time $\tau$ ($I(\tilde T_i\leq \tau)=1$ and $\Delta_i=1$), or (2) remain event-free at time $\tau$ and have not been censored before time $\tau$ ($I(\tilde T_i>\tau)=1$). The weights are the inverse of the estimated $G(\tilde T_i)$ (for people with $I(\tilde T_i\leq \tau)=1$ and $\Delta=1$) or $G(\tau)$ (for people with $I(\tilde T_i> \tau)=1$), assuming independent censoring. The regression excludes individuals who are censored before time $\tau$. After fitting the weighted non-negative least squares regression, the estimated coefficients are scaled to sum to 1, as outlined above. The resulting SL estimator for the conditional survival probability based on $L_{IPCW}$ is 
\begin{equation}
    S^{SL}(\tau|X=x_i)=\sum_{j=1}^{p}\hat \alpha_j \phi_{i,j}.
    \label{eq:SL.est.ipcw}
\end{equation}
The regression in equation (\ref{eq:SL.ensemble.IPCW}) is fitted using a data set with one row per individual, because $\phi_{j}$ refers to the survival probability at the specified time $\tau$ only. This differs from the fitting of the regressions based on $L_{L2}$ and $L_{loglik}$, which are fitted on the long data.

\section{Super learners for continuous time-to-event data}
\label{sec:continuous}

\subsection{Overview}
\label{sec:continuous.overview}

The discrete-time SL has some disadvantages. The discretisation of follow-up time means that we do not make full use of the information on event times that are in the data. Flexible modelling of time as a covariate in candidate learners for the baseline hazard is also a challenge. The two-step approach recommended by \citet{polley_superlearner_2011} is not implemented in the main R package \texttt{SuperLearner} available for implementation (see Section \ref{sec:implementation}). The requirement to discretise the data, resulting in a data set with many rows per individual, can also bring computational challenges.

In this section we describe two continuous-time SL approaches, i.e. approaches that do not require discretisation of follow-up time. \citet{westling_inference_2023} described a SL for continuous-time data (or a mixture of continuous- and discrete-time data). They are motivated by the need to estimate conditional survival and censoring distributions, $S(t|X)$ and $G(t|X)$, because they feature as nuisance functions in a wider causal estimation procedure aiming to estimate survival curves under different treatments using a doubly-robust approach. We focus on their SL for estimation of target estimand $S(t|X)$. Another continuous-time SL approach was proposed by \citet{munch_jossl_2025}, who referred to their approach as the `joint survival super learner' (in an earlier version of their paper it was referred to as the `state learner' \cite{munch_statelearner_2024}). They focus on use of a SL to estimate state occupation probabilities in a setting that can include more than one event type of interest, for example where there are competing events. We focus on the use of their approach when there is just a single absorbing event type of interest (e.g. death). The method of \citet{westling_inference_2023} includes an ensemble step, whereas the joint survival SL of \citet{munch_jossl_2025} is a non-ensemble SL. Both methods involve loss functions that require estimates of the conditional censoring distribution $G(t|X)$, but they enable this to be done in a flexible way, making use of a SL for the conditional censoring distribution. 

\subsection{Continuous-time super learner: Westling et al. (2023)}

\subsubsection{SL steps 1-3: Cross-validation}
\label{sec:westling.setup}

This method targets estimation of the whole conditional survival distribution $S(t|X)$, for $0<t\leq \tau$ say. Any model that gives rise to an estimate of $S(t|X)$ for any value of $t$ (within a specified range) could be a candidate learner for inclusion in the SL. In their example application \citet{westling_inference_2023} include the following candidate learners: parametric survival models, semiparametric proportional hazard models, generalized additive Cox models, and random survival forests. Each candidate learner $j=1,\ldots,p$ is fitted in training set $D_{-k}$ and is then used to obtain estimates $\hat S_j(t|X=x_i)$ for individuals in validation set $D_k$. In order to evaluate predictive performance of candidate learners (see step 4), this approach requires estimates $\hat S_j(t|X=x_i)$ for each $t$ on a grid of times. Suppose that we divide the time interval of interest $(0,\tau)$ into a grid $V$ of evenly spaced times with differences $v$ (i.e. $V=\{v,2v,3v,\ldots,\tau-v,\tau\}$). Cross-validated estimates of $\hat S_j(t|X=x_i)$ are obtained at each time $t$ in $V$, i.e. $\hat S_j(v|X=x_i),\hat S_j(2v|X=x_i),\ldots,\hat S_j(\tau|X=x_i)$. Using each candidate learner $j$ we can then create a vector of estimated conditional survival probabilities for each person $i$ in $D$ at each time point on the grid $V$: $\hat S_{i,j}^{V}=\left(\hat S_j(v|X=x_i),\hat S_j(2v|X=x_i),\ldots,\hat S_j(\tau|X=x_i)\right)$, for $i=1,\ldots,n$ and $j=1,\ldots,p$. We emphasise that the use of the time grid here is different from the discretisation of time for estimation of $S(\tau|X)$ in the discrete-time SL. Using a fine time grid is recommended here, as the grid is used in the next step to estimate an integral.  

\subsubsection{SL Step 4: Estimating the mean expected loss}
\label{sec:westling.loss}

The loss function proposed by \citet{westling_inference_2023} for the conditional survival distribution over the interval $(0,\tau)$, denoted $L_{S,G}$, is
\begin{equation}
    L_{S,G}=\int_{0}^{\tau}\left(S(t|x)-f_G(t)\right)^2dt
        \label{eq:Westling.S.loss}
\end{equation}
where $f_G(t)=1-\Delta I(\tilde{T}\leq t)/G(\tilde{T}|X)$. This loss function is a weighted integrated Brier score. It can be used under a pre-specified model for the censoring distribution $G(t|X)$, or $G(t|X)$ can itself be estimated using a SL using an approach that combines SLs for both $S(t|X)$ and $G(t|X)$. We first outline the use of the approach of \citet{westling_inference_2023} under a pre-specified model for $G(t|X)$, before outlining their extension to using SLs for both $S(t|X)$ and $G(t|X)$ (Section \ref{sec:westling.cens}). Suppose that a model for the conditional survival distribution $G(t|X)$ has been pre-specified, for example as a Cox model or random survival forest, and let $\hat G(\tilde{T}|X)$ denote the resulting cross-validated estimates. Based on $L_{S,G}$, the expected loss for candidate learner $j$ in fold $k$ is obtained as follows, where the integral in (\ref{eq:Westling.S.loss}) is approximated via a sum using the grid of times $V$:
\begin{equation}
    \hat{L}_{S,G}^{(j,k)}=\sum_{t \in V}\frac{v}{|D_k|}\sum_{i\in D_k}\left\{\hat S_j(t|X=x_i)-\left(1-\frac{\Delta_i I(\tilde{T}_i\leq t)}{\hat G(\tilde{T}|X=x_i)} \right)\right\}^2.
    \label{eq:Westling.S.loss.est}
\end{equation}
As before, we then obtain the estimated mean expected loss $\bar{L}^{(j)}$ for candidate learner $j$ ($j=1,\ldots,p$) using the average of $\hat{L}_{S,G}^{(j,k)}$ over the $K$ folds (step 4 in Section \ref{sec:overall.SL.steps}). As before, the best performing individual learner (the one with smallest mean expected loss) is the non-ensemble SL.

\subsubsection{SL step 5: Deriving an ensemble}
\label{sec:westling.ensemble}

Derivation of the SL ensemble makes use of the cross-validated estimated survival probabilities on the time grid $V$ for the $p$ candidate learners, $\hat S_{i,j}^{V}$ ($j=1,\ldots,p$). The aim is to derive a weighted combination of the estimates from the $p$ candidate learners such that the loss function $L_{S,G}$ in (\ref{eq:Westling.S.loss}) is minimised. The loss function involves $f_G(t)=1-\Delta I(\tilde{T}\leq t)/G(\tilde{T}|X)$ and we can obtain the estimates $\hat f_G(i,t)=\left(1-\frac{\Delta_i I(\tilde{T}_i\leq t)}{\hat G(\tilde{T}_i|X=x_i)} \right)$ for each person at each time point $t$ in the grid $V$, giving the outcome vector $\hat f^V_G(i)=\left(\hat f_G(i,v),\hat f_G(i,2v),\ldots,\hat f_G(i,\tau)\right)$. The ensemble SL is then derived by performing a non-negative least squares linear regression of $\hat f^V_G(i)$ on $\hat S_{i,1}^{V},\ldots,\hat S_{i,p}^{V}$ across all times in $V$ combined, i.e. a model with linear predictor $\sum_{j=1}^{p}\alpha_j \hat S_{j}(t|X=x_i)$, where the $\alpha_j$ are constrained to be non-negative. The model is fitted on a `long' data set which includes the estimates $\hat f_G(i,t)$ and $\hat S_{j}(t|X=x_i)$ for each person $i$ in $D$ at each time $t$ on the grid $V$. After fitting the model the estimates of $\alpha_1,\ldots,\alpha_{p}$ are scaled to sum to 1 as described earlier. The resulting SL estimator is of the form
\begin{equation}
    S^{SL}(t|X)=\sum_{j=1}^{p}\hat \alpha_j \hat S_{j}(t|X).
    \label{eq:SL.est.westling}
\end{equation}
As before, to obtain an estimate for a new individual, the  candidate learners are refitted on the full data for use in (\ref{eq:SL.est.westling}).

\subsubsection{Extending to estimate the censoring distribution via super learner}
\label{sec:westling.cens}

We now outline how this method extends to allow for using a SL for the censoring distribution $G(t|X)$ as well as for $S(t|X)$. This requires candidate learners and loss functions for both distributions. \citet{westling_inference_2023} specify the following loss function for the conditional censoring distribution over the interval $(0,\tau)$, which is analogous to the loss function $L_{S,G}$ for the survival distribution in (\ref{eq:Westling.S.loss}):
\begin{equation}
    M_{G,S}=\int_{0}^{\tau}\left\{G(t|x)-f_S(t)\right\}^2dt
        \label{eq:Westling.G.loss}
\end{equation}
where $f_S(t)=1-(1-\Delta)I(\tilde{T}< t)/S(\tilde{T}|X)$. For a given candidate learner for the censoring distribution, the  expected loss based on $M_{G,S}$ can be obtained using a similar formula to that in (\ref{eq:Westling.S.loss.est}). The loss function for the survival distribution depends on the censoring distribution, and vice versa. 

To derive SLs for both $G(t|X)$ and $S(t|X)$, \citet{westling_inference_2023} proposed an iterative approach. We assume $p$ candidate learners for $S(t|X)$ and $q$ candidate learners for $G(t|X)$. The steps are as follows:
\begin{enumerate}
    \item Fit candidate learners for both $S(t|X)$ and $G(t|X)$ in a cross-validated way based on $K$ data folds, as outlined in SL steps 1-3 for $S(t|X)$ (Section \ref{sec:overall.SL.steps}). Obtain predicted probabilities from each model for each person on the time grid $V$. This gives $\hat S_{i,j}^{V}$, as introduced in Section \ref{sec:westling.setup}, for $i=1,\ldots,n$; $j=1,\ldots,p$. The corresponding estimates for the censoring distribution  are $\hat G_{i,j^{\prime}}^{V}=\left(\hat G_{j^{\prime}}(v|X=x_i),\hat G_{j^{\prime}}(2v|X=x_i),\ldots,\hat G_{j^{\prime}}(\tau|X=x_i)\right)$ for $i=1,\ldots,n$; $j^{\prime}=1,\ldots,q$.
    
    \item In the full data $D$ fit a single model to enable estimation of $\hat G(t|X)$. This can be any model and it just provides us with an initial estimate. Use this model to obtain an estimate of the probability of remaining uncensored, $\hat G(\tilde{T}_i|X=x_i)$, for each individual at their own observed event or censoring time $\tilde{T}_i$. 
    
    \item Calculate $\hat f_G(i,t)=\left(1-\frac{\Delta_i I(\tilde{T}_i\leq t)}{\hat G(\tilde{T}_i|X=x_i)} \right)$ for each person at each time point $t$ in the grid $V$ using the current estimator of the censoring distribution $G(t|X)$. Using the current values of $\hat f^V_G(i)=\left(\hat f_G(i,v),\hat f_G(i,2v),\ldots,\hat f_G(i,\tau)\right)$ fit a non-negative least squares regression of $\hat f^V_G(i)$ on $\hat S_{i,1}^{V},\ldots,\hat S_{i,p}^{V}$ and scale the coefficients to sum to 1, as outlined in Section \ref{sec:westling.ensemble}. This gives the current SL estimator for $S(t|X)$, $S^{SL}(t|X)=\sum_{j=1}^{p}\hat \alpha_j \hat S_{j}(t|X)$.
    
    The following steps are then iterated:
    
    \item  Calculate $\hat f_S(i,t)=\left(1-\frac{(1-\Delta_i) I(\tilde{T}< t)}{\hat S(\tilde{T}|X=x_i)} \right)$ for each person at each time point $t$ in the grid $V$ using $\hat S^{SL}(t|X)$ from the previous step. Fit a non-negative least squares regression of $\hat f^V_S(i)=\left(\hat f_S(i,v),\hat f_S(i,2v),\ldots,\hat f_S(i,\tau)\right)$ on $G_{i,1}^{V},\ldots,G_{i,q}^{V}$ and scale the coefficients to sum to 1. This gives the current SL estimator for $G(t|X)$, $G^{SL}(t|X)=\sum_{j^{\prime}=1}^{q}\hat \beta_j \hat G_{j^{\prime}}(t|X)$.
    
    \item Calculate $\hat f_G(i)$ using $\hat G^{SL}(t|X)$ from the previous step. Fit a non-negative least squares regression of $\hat f^V_G(i)$ on $\hat S_{i,1}^{V},\ldots,\hat S_{i,p}^{V}$ and scale the coefficients to sum to 1, to give an updated SL estimator $S^{SL}(t|X)$.
    
    \item Obtain estimates of $S(t|X)$ for all individuals for all time points in the grid $V$ based on the current $S^{SL}(t|X)$, giving $\hat S^{SL}(t|X=x_i),i=1,\ldots,n$. These estimates are now compared these with the estimates $\hat S^{SL}(t|X=x_i)$ from the previous iteration. Calculate the maximum absolute difference across all $n\times |V|$ person-time combinations, where $|V|$ is the number of time in the grid $V$. If the maximum absolute difference is less than a small pre-determined value $\epsilon$ then we stop. \citet{westling_inference_2023} use $\epsilon=10^{-5}$ in the implementation in their code. Otherwise we return to step 4 for another iteration, repeating steps 4-6 until the maximum absolute difference is less than $\epsilon$. 
\end{enumerate}

\citet{westling_inference_2023} point out that their iterative approach only requires the cross-validated fitting of the candidate learners to be performed once (in step 1), with the only refitting being of the weighted regressions used to derive the ensemble, and this part is not computationally expensive.

As in the approach of \citet{polley_superlearner_2011} all candidate learners with a non-zero weight in the final SL ensemble are refitted on the full data. To obtain a prediction of $S(t|X)$ for a new individual we use these estimators combined with the cross-validated coefficients obtained from the above iterated process. 

\subsection{The joint survival super learner: Munch \& Gerds (2025)}
\label{sec:munch}

\subsubsection{SL steps 1-3: cross-validation}
\label{sec:Munch.setup}

The joint survival SL approach of \citet{munch_jossl_2025} makes use of state occupation functions. In our setting of a single event of interest, at any time $t$ an individual in $D$ can be considered to be in one of three possible states denoted $s$: (i) they remain uncensored and have not had the event ($s=0$), (ii) they have been observed to have had the event ($s=1$), or (iii) they have been censored ($s=-1$). Let $\Lambda(t|X)$ denote the conditional cumulative hazard for the event of interest, and let $\Gamma(t|X)$ denote the conditional cumulative hazard for censoring. The conditional probabilities of being in state $s=0,1,-1$ at time $t$ (in other words, the state occupation probabilities, the risks, or the cumulative incidences) can be written as follows:
\begin{equation}
     F(t,0,X)=\Pr(\tilde T>t|X)=\exp\left(-\Lambda(t|X)-\Gamma(t|X)\right),
\end{equation}
\begin{equation}
    F(t,1,X)=\Pr(\tilde{T}\leq t,\Delta=1|X)=\int_{0}^{t}F(s^{-},0|X)\Lambda(ds|X),
\end{equation}
\begin{equation}
    F(t,-1,X)=\Pr(\tilde{T}\leq t,\Delta=0|X)=\int_{0}^{t}F(s^{-},0|X)\Gamma(ds|X).
\end{equation}
We note that $F(t,1,X)$ differs from 1 minus the survival probability, $1-S(t|X)=\Pr(T\leq t|X)$, because $F(t,1,X)$ is a probability of being \emph{observed} to have had the event up to $t$ in the world in which censoring occurs. 

The joint survival SL involves specifying candidate learners for the cumulative hazards of both the event and the censoring, $\Lambda(t|X)$ and $\Gamma(t|X)$. We assume $p$ candidate learners for the event ($\Lambda(t|X)$) and $q$ candidate learners for censoring ($\Gamma(t|X)$). The candidate learners could be any method for continuous time-to-event outcomes that can output a conditional cumulative hazard. \citet{munch_jossl_2025} consider Cox regression, the Nelson-Aalen estimator, a Cox Lasso, and a random survival forest. 

As we will see in Section \ref{sec:munch.loss}, the loss function is in the form of an integrated Brier score for all three states combined, and depends on the state occupation probabilities $F(t,s,X)$ ($s=-1,0,1$). To estimate the mean expected loss we need to obtain cross-validated estimates of $F(t,s,X=x_i)$ ($s=-1,0,1$) for each individual in $D$. The $F(t,s,X)$ ($s=-1,0,1$) involve integrals and these can be estimated based on a given pair of candidate learners (for $\Lambda(t|X)$ and $\Gamma(t|X)$) using a fine time grid. As for the method of \citet{westling_inference_2023} we assume a grid $V=\{v,2v,3v,\ldots,\tau-v,\tau\}$ of evenly spaced times with differences $v$. The practical steps for obtaining cross-validated estimates of $F(t,s,X)$ ($s=-1,0,1$) are:
\begin{enumerate}
    \item Divide the data $D$ into $K$ folds, $k=1,\ldots,K$, as outlined in step 1 of the general SL steps given in Section \ref{sec:overall.SL.steps}.
    \item Fit each of the $p$ candidate learners for $\Lambda(t|X)$ and each of the $q$ candidate learners for $\Gamma(t|X)$ to the training set $D_{-k}$, for $k=1,\ldots,K$.
    \item Using candidate learner $j$ ($j=1,\ldots,p$) fitted in training set $D_{-k}$, obtain estimates of the conditional cumulative hazard for the event $\Lambda(t|X)$ in validation set $D_k$ on a grid of times $V$, with the resulting estimates at time $t$ denoted $\Lambda_j(t|X=x_i)$ ($t\in V; i\in D_k$). 
    \item Using candidate learner $j^{\prime}$ ($j^{\prime}=1,\ldots,q$) fitted in training set $D_{-k}$, obtain estimates of the conditional cumulative hazard for censoring $\Gamma(t|X)$ in validation set $D_k$ on a grid of times $V$, with the resulting estimates at time $t$ denoted $\Gamma_{j^{\prime}}(t|X=x_i)$ ($t\in V; i\in D_k$). 
    \item For individuals in validation set $D_k$ obtain estimates of the state occupation probabilities $F(t,s,X),s=-1,0,1$ for each pair of candidate learners $(j,j^{\prime})$ fitted in training set $D_{-k}$:
    $$
    \hat{F}_{j,j^{\prime}}(t,0,x_i)=\exp\left(-\hat \Lambda_{j}(t|X=x_i)-\hat\Gamma_{j^{\prime}}(t|X=x_i)\right),
    $$
    $$
    \hat{F}_{j,j^{\prime}}(t,1,x_i)=\sum_{l\leq t}\hat{F}_{j,j^{\prime}}(l-1,0,x_i)\left(\hat{\Lambda}_{j}(l|X=x_i)-\hat{\Lambda}_{j}(l-1|X=x_i)\right),
    $$
    $$
    \hat{F}_{j,j^{\prime}}(t,-1,x_i)=\sum_{l\leq t}\hat{F}_{j,j^{\prime}}(l-1,0,x_i)\left(\hat{\Gamma}_{j^{\prime}}(l|X=x_i)-\hat{\Gamma}_{j^{\prime}}(l-1|X=x_i)\right).
    $$
    \item Repeat steps 3-5 for each $k=1,\ldots,K$, giving cross-validated estimates of $F(t,s,X),s=-1,0,1$ for all individuals in $D$.
\end{enumerate}

\subsubsection{SL step 4: Loss function}
\label{sec:munch.loss}

\citet{munch_jossl_2025} proposed using an integrated Brier score as the loss function, based on a version of the Brier score that considers both the event and censoring processes. In this setting the Brier score over all three states at time $t$ is:
\begin{equation}
 B(t)=\sum_{s=-1,0,1}\left(F(t,s,X)-I(\eta(t)=s)\right)^2   
\end{equation}
where $\eta(t)$ denotes the observed state ($s=-1,0,1$) that a person is in at time $t$. By its definition, $\eta(t)$ is observed at any time for all individuals, hence we do not need to use censoring weights to estimate this Brier score. The integrated Brier score over the interval $(0,\tau)$ is then $IBS(\tau)=\int_{0}^{\tau}B(t) dt$. 

The estimated Brier score at time $t$ based on candidate learner pair $(j,j^{\prime})$ for the event and censoring cumulative hazards $\Lambda(t|X)$ and $\Gamma(t|X)$ applied to validation set $D_k$, is
\begin{equation}
     \hat{B}^{(j,j^{\prime},k)}(t)=\sum_{s=-1,0,1}\frac{1}{|D_k|}\sum_{i\in D_k}\left(\hat F_{j,j^{\prime}}(t,s,x_i)-I(\eta_i(t)=s)\right)^2.
\end{equation}
This can be calculated for each $t$ in the time grid $V$ and the corresponding integrated Brier score is estimated using
\begin{equation}
    \widehat{IBS}^{(j,j^{\prime},k)}(\tau)=\sum_{t\in V }v\times \hat{B}^{(j,j^{\prime},k)}(t).
    \label{eq:munch.ibs.est}
\end{equation}
The estimated mean expected loss for the pair of candidate learners $(j,j^{\prime})$ is $\frac{1}{K}\sum_{k=1}^{K}\widehat{IBS}^{(j,j^{\prime},k)}(\tau)$. 

\subsubsection{The non-ensemble super learner}
\label{sec:munch.step5}

The approach of \citet{munch_jossl_2025} is designed to derive a non-ensemble SL. The final joint survival SL is given by the pair of candidate learners $(j,j^{\prime})$ with the lowest cross-validated mean expected loss estimate based on (\ref{eq:munch.ibs.est}). In our setting where the target estimand is $S(\tau|X=x)=\Pr(T>\tau|X=x)$ the SL estimator is given by $S^{SL}(\tau|X)=\exp(-\hat \Lambda_j (\tau|X))$, where $j$ is the selected candidate learner for $\Lambda (t|X)$ and the estimates $\hat \Lambda_j (\tau|X)$ are obtained based on fitting candidate learner $j$ to the full data $D$. The selected candidate learner for the censoring distribution is not used in the estimator $S^{SL}(\tau|X)$, but the selection of candidate learner $j$ for $\Lambda (t|X)$ was based on considering pairs of candidate learners for $\Lambda(t|X)$ and $\Gamma(t|X)$.

\section{Super learner implementation in R}
\label{sec:implementation}

\subsection{Discrete-time super learner}

The discrete-time SL approach of \citet{polley_superlearner_2011} can be implemented in R using the \texttt{SuperLearner} package \citep{polley_package_2024}. This package implements SL for binary or continuous outcomes and is not specifically designed for discretised time-to-event outcomes. The discretisation of the data needs to be done before using the \texttt{SuperLearner} function. The discretisation can be done using time periods of equal length or by dividing time up on the basis of quantiles of the event times, for example using the \texttt{survsplit} function in the \texttt{survival} package. There is little formal advice on this aspect, but in general discretising time can result in biased estimates of risk if the discretisation is too crude. In their example, \citet{polley_superlearner_2011} discretise time using 30 time intervals defined by quantiles of the event times. 

There is a wide choice of candidate learners available in the \texttt{SuperLearner} package. It includes a built in library of 40 learners. Users can also specify \emph{screening algorithms}, which perform variable selection using specified methods (e.g. Lasso) before then implementing the selected candidate learners for the selected variables. The \texttt{SuperLearner} package includes 8 built-in screening algorithm options. The package also allows users to specify their own learners and screening algorithms. The github repository \url{https://github.com/ecpolley/SuperLearnerExtra} provides additional options. 

Three loss functions for use in the discrete-time SL approach were discussed in Section \ref{sec:discrete.loss}. The \texttt{SuperLearner} function implements the L2 loss function $L_{L2}$ (using \texttt{method = "method.NNLS"}) and the log-likelihood loss function $L_{loglik}$ (using \texttt{method = "method.NNloglik"}), but not the IPCW loss function $L_{IPCW}$, despite $L_{IPCW}$ being recommended for use when the target estimand is a survival probability \citep{polley_superlearner_2011}. However, previous work has found results obtained using the three loss functions to be very similar \citep{tanner_dynamic_2020}. A number of other loss functions are available in the package, but these would not necessarily be suitable for discrete-time-to-event outcomes. The \texttt{SuperLearner} package does not directly implement the 2-stage approach to allowing flexibility in modelling the baseline hazard using a GAM. 

The use of cross-validation is a fundamental component of the SL used to derive the mean expected loss (also called the cross-validated risk), and is the basis for the oracle property. Because the cross-validation involves randomly dividing the data into $K$ folds, we should set the random number seed in order to obtain reproducible results. It can also be of interest to obtain a cross-validated risk for the SL ensemble itself as well as for each of the component candidate learners. This can be done using the \texttt{cv.SuperLearner} function, which repeats the whole SL procedure in a cross-validated way - i.e. the SL procedure is repeated in a series of training sets and the loss function for the SL is then estimated in a series of validation sets (folds). Nested within this is the cross-validation that is part of each SL.

To apply the SL to obtain survival probability estimates on an external data set, the fits of the candidate learners on the full training data are combined with the cross-validated coefficients to form the SL ensemble. The \texttt{SuperLearner} package \citep{polley_package_2024} does this by allowing the user to provide training and test data sets.

\subsection{Continuous-time super learner}

The iterative SL approach of \citet{westling_inference_2023} outlined in Section \ref{sec:westling.cens} can be implemented using their \texttt{survSuperLearner} package, which can be installed from Github (\url{https://github.com/tedwestling/survSuperLearner}). This package accommodates a number of candidate learners: Kaplan-Meier estimator, Cox regression, parametric survival models (exponential, Weibull, log-logistic, piecewise constant hazard), generalised additive Cox model, random survival forest, and an approach that estimates a piecewise constant hazard using a separate SL for a binary outcome (implemented using the \texttt{SuperLearner} package). The number of folds to be used in the cross-validation can be specified as an option in the function. Each of the candidate learners can be combined with different screening algorithms. The screening algorithms included are: include all covariates; include covariates selected using a Cox Lasso regression; include covariates based on separate univariable Cox regressions for each covariate (with selection based on the Wald-test p-value being below a specified threshold). For the Kaplan-Meier method no covariates are used and the same results are obtained under all screening algorithms. The code also includes a template for writing your own prediction or screening algorithm. The \texttt{survSuperLearner} package also allows for additional observation weights.  

The joint survival SL of \citet{munch_jossl_2025} can be implemented using R code from \url{https://github.com/amnudn/joint-survival-super-learner} (\texttt{jossl}). The code currently accommodates up to two event types plus censoring, and it accommodates the following set of candidate learners for the cumulative hazards: Cox regression, the Nelson-Aalen estimator, a Cox Lasso, and a random survival forest. The \texttt{jossl} code performs 5-fold cross-validation.

The continuous-time SLs use a time grid to estimate certain integrals (denoted $V$ in Section \ref{sec:continuous}). The grid can be specified in both \texttt{survSuperLearner} and \texttt{jossl}. The default in \texttt{survSuperLearner} is to use 250 time points equally spaced between 0 and the last uncensored follow-up time (for the event survival distribution) or the last censored follow-up time minus a small number (for the censoring distribution). In \texttt{jossl} the default is 100 equally spaced time points from 0 up to the time horizon of interest.

The \texttt{survSuperLearner} package enables predictions to be obtained on an external data set by allowing the user to provide training and test data sets, as in \texttt{SuperLearner}.

\section{Application to the Rotterdam data}
\label{sec:application}

\subsection{Data}

We illustrate the use of the three SL approaches described in this paper through an application to predict conditional survival probabilities following breast cancer diagnosis using the `Rotterdam' data set, which is freely available as part of the \texttt{survival} package in R \citep{therneau_survival_2024}. This data set has been widely used to illustrate survival analysis methods \citep{royston_external_2013}. It includes data on individuals who underwent surgery for primary breast cancer between 1978 and 1993, and whose data were recorded in the Rotterdam Tumour Bank. Individuals were followed up for disease recurrence and death for up to a maximum of 19.3 years. The data include information on the following potential predictor variables: year of diagnosis, age, menopausal status (0: premenopausal, 1: postmenopausal), tumour size ($\leq 20$mm, $20-50$mm, $>50$mm), differentiation grade (2 or 3), number of positive lymph nodes, progesterone receptors (continuous, fmol/l), estrogen receptors (continuous, fmol/l), chemotherapy (0: no, 1: yes), hormonal treatment (0: no, 1: yes). 

\subsection{Methods: implementation of the super learners}

The target estimand is the conditional survival probability up to 10 years. We illustrate the use of the methods in the setting in which a training data set is available for model development, and it is of interest to apply the resulting prediction algorithm in an external test data set. We divide the Rotterdam data randomly into a sample of 2087 individuals (70\%) used for the development/training, with the remaining 895 individuals (30\%) being used as a test set in which we will apply the SL and also evaluate its performance externally. We emphasise that using the SL involves cross-validation within the training data.

To implement the methods we make use of the packages and code referenced in Section \ref{sec:implementation}. Code for reproducing the analyses and results is provided at  \url{https://github.com/ruthkeogh/superlearner_survival_tutorial}. We also provide `by hand' implementations (i.e. not using the packages) of the discrete-time SL method and the joint survival SL approach to clearly illustrate the steps. We first use a similar simple set of candidate learners across the three SL methods, though they way they are used differs slightly across methods. We then implement a richer set of candidate learners where possible using the available packages. For all SLs we used 5-fold cross-validation.

For the discrete-time SL of \citet{polley_superlearner_2011} we divide time up into 10 or 100 periods of equal length to illustrate results with different levels of discretisation. When using 10 time periods we enter time $t$ as a categorical variable for each period, and when using 100 time periods we enter $t, t^2, t^3$ as continuous variables. For the implementation using a simple set of learners, the candidate learners used are: a non-parametric empirical estimate including no covariates, logistic regression with and without Lasso selection of covariates, and a random forest (using the default settings of \texttt{SL.ranger}). We perform the analysis using the $L2$ and $loglik$ loss functions. The IPCW loss function is not accommodated in the \texttt{SuperLearner} package, but we illustrate its use in our by-hand implementation. For the discrete-time SL using a richer set of candidate learners we include random forests, GAMs and XGboost with different sets of tuning parameters, alongside the simple set of learners, following the set used by \citet{tanner_dynamic_2020} (see Supplementary Table \ref{table:ensemble.weights.disc.rich}). 

For the continuous-time SL of \citet{westling_inference_2023} we use the iterative approach in which a SL is used for both the event and censoring distributions, as is implemented in the \texttt{survSuperLearner} package. In the iterative process we use a value of $\epsilon=10^{-5}$ as the convergence threshold, as applied in the available code. For the implementation using a simple set of learners, the candidate learners used are the Kaplan-Meier estimator, Cox regression with and without Lasso selection of covariates, and a random survival forest (using the default settings in R, which includes to use a miniumum terminal node size of 15). We also implement this SL using the full set of candidate learners that are built into the \texttt{survSuperLearner} package, which additionally include exponential, Weibull and log-logistic models, Cox regression including GAMs, and piecewise constant hazard regressions (see Supplementary Table \ref{table:ensemble.weights.cont.rich}). 

For the continuous-time SL (`joint survival SL') of \citet{munch_jossl_2025} we use the same set of candidate learners for the event and censoring distributions. These are the Nelson-Aalen estimator, Cox regression with and without Lasso selection of covariates, and random survival forest. This gives 16 possible pairs of candidate learners for the event and censoring distributions.

The continuous-time SLs use a time grid to estimate certain integrals, and we use the defaults for this. In the by-hand implementation we evaluate the integrals using a time grid defined by all observed event times. 

Using the continuous-time data we also apply each of the simple candidate learners (Cox regression, Cox regression with Lasso selection of covariates, random survival forest) separately, for comparison with the SLs.

\subsection{Methods: application and evaluation in the test data}

To evaluate the performance of the SLs in the test dataset we use a range of measures of discrimination, calibration, and overall performance, guided by the recommendations of \citet{mclernon_assessing_2022}. We estimate the Brier score at time 10 years and the scaled Brier score, which compares the estimated Brier score from the model to the Brier score obtained if everyone is assigned the same survival probability obtained from a Kaplan-Meier analysis \citep{graf_assessment_1999}. 
To assess discrimination we estimate the C-index up to time 10 years, and the cumulative dynamic AUCt at time 10 years. For the C-index we use Uno's C \citep{uno_c-statistics_2011} which accommodates censoring through weights. Our code also shows how to obtain calibration plots, but we do not show these here. 

All of the measures used to evaluate the predictive performance require inverse probability of censoring weights, which are estimated using a model for the censoring distribution fitted in the test data. For this it is commonly assumed that censoring is independent, with censoring weights being estimated as the inverse of the Kaplan-Meier estimates for the censoring survival probabilities. We apply this approach. In the example code we also illustrate estimation of the censoring weights using a SL for the censoring distribution, fitted in the test data. 

\citet{munch_jossl_2025} compared SL methods in a simulation study using the scaled Brier score. However, the scaled Brier score was obtained for a large data set without any censoring, hence they avoided having to deal with censoring when estimating the Brier score.

\subsection{Results}

Table \ref{table:prediction.measures} shows the estimated measures of predictive performance using the individual (continuous-time) learners, the discrete-time ensemble SL, the continuous-time ensemble SL, and the continuous-time non-ensemble SL. Supplementary Tables \ref{table:ensemble.weights.disc.simple}-\ref{table:ensemble.weights.cont.rich} show the coefficients (or weights) attached to different candidate learners in the discrete-time and continuous-time ensemble SLs, based on SLs using simple and richer sets of candidate learners.

Of the individual learners the random survival forest gives the best predictive performance, in terms of having the lowest Brier score, highest scaled Brier score, and the best discrimination. The standard Cox regression has the worst predictive performance overall.

In the discrete-time SL, the version in which we discretised using 100 periods has worse overall performance, as measured by the Brier score, compared to the version in which we discretised using only 10 time periods. However the discrimination measures are very similar using the two levels of time discretisation. The difference is likely to be in part due to poor modelling of the baseline hazard in the version using 100 time periods. Recall that when using 10 time periods we include an indicator for each time period in each candidate learner, whereas when using 100 time periods we include linear, quadratic and cubic terms for time. The discrete-time SL using 100 time periods has worse overall performance than using the Cox model as an individual learner, but better discrimination.

The continuous-time SLs give better performance than the discrete-time SLs, in particular when looking at overall performance measured using the Brier score. The continuous-time non-ensemble SL of \citet{munch_jossl_2025} selects the random survival forest as the best model for the event time distribution, and hence the results are the same as shown when using the random survival forest as an individual learner. The continuous-time ensemble SL of \citet{westling_inference_2023} gives almost identical predictive performance to the continuous-time non-ensemble SL (and the random survival forest). 

Using a richer set of candidate learners gives better predictive performance, though the gains are modest in this example (Table \ref{table:prediction.measures}). When implemented using a simple set of candidate learners, the discrete-time SL with 10 time periods places most weight on the GLM with Lasso and on the random forest using both the $L2$ and $loglik$ loss functions (Supplementary Table \ref{table:ensemble.weights.disc.simple}). The GLM with Lasso was the single best performing model using both loss functions. The version using the $L2$ loss function places zero weight on the standard GLM, whereas the version using the $loglik$ loss function gives that model a small weight. By contrast, the discrete-time SL with 100 time periods and using a simple set of candidate learners places most weight on the GLM and zero weight on the GLM with Lasso, for both loss functions. With 100 time periods and using the $L2$ loss function, the single best performing model is the GLM with Lasso, however the standard GLM has very similar performance, which can explain why the GLM with Lasso was assigned zero weight in the ensemble. With 100 time periods and using the $loglik$ loss function, the single best performing model was the standard GLM. When we use a richer set of candidate learners (Supplementary Table \ref{table:ensemble.weights.disc.rich}), the discrete-time SL with 10 time periods places most weight on a GAM with 3 degrees of freedom, but also places non-zero weight on the mean estimator, the GLM with Lasso, a GAM with 2 degrees of freedom, one XGBoost learner (out of 7), and one random forest (out of 3). 

The continuous-time ensemble SL of \citet{westling_inference_2023} using the simpler set of candidate learners assigns the most weight to the random survival forest for the event distribution (Supplementary Table \ref{table:ensemble.weights.cont.simple}), some weight to the Cox model with Lasso, and zero weight to the Kaplan-Meier or the standard Cox model. When using a richer set of learners (Supplementary Table \ref{table:ensemble.weights.cont.rich}), all of the weight is shared between the Cox regression with continuous covariates modelled using GAMs and the random survival forest. Corresponding results for the censoring distribution are shown in Supplementary Tables \ref{table:ensemble.weights.cont.simple} and \ref{table:ensemble.weights.cont.rich}.

Implementations of the discrete-time approach and the continuous-time approach of \citet{munch_jossl_2025} performed `by hand' using a simple set of candidate learners give similar results as found using the packages (see Supplementary Table \ref{table:prediction.measures.byhand}). 

\begin{table}[!ht]
    \caption{Results from the application of individual learners and SLs to the Rotterdam data: measures of predictive performance in the test data set.}
    \label{table:prediction.measures}
    \centering
    \begin{tabular}{lllll}
    \hline
    Prediction method&Brier score&Scaled Brier&C-index&AUCt\\
    \hline
    \multicolumn{5}{l}{\bf Individual learners}\\
    Cox regression  &0.211 &14.5  &70.3 &71.7 \\
    Cox regression - Lasso  &0.209 &15.3  &71.0 &74.7 \\
    Random survival forest  &0.196 &20.6  &71.7 &75.8 \\
    &&&&\\
    \multicolumn{5}{l}{\bf Super learners: simple set of learners (see Supplementary Tables \ref{table:ensemble.weights.disc.simple} and \ref{table:ensemble.weights.cont.simple})}\\
    Discrete-time SL: 10 periods, $L2$ loss  &0.204 &17.2  &72.0 &74.6 \\
    Discrete-time SL: 10 periods, $loglik$ loss  & 0.203& 17.6 &71.9 & 74.4\\
    &&&&\\
    Discrete-time SL: 100 periods, $L2$ loss  &0.212 &13.9  &71.9 &75.3 \\
    Discrete-time SL: 100 periods, $loglik$ loss &0.213 &13.5  &71.6 &74.8 \\
    &&&&\\
    Continuous-time SL: Westling et al. 2023&0.196 &20.6  &72.0 &75.8 \\
    Continuous-time SL: Munch \& Gerds 2025  &0.196 &20.6  &71.7 &75.8 \\
    &&&&\\
    \multicolumn{5}{l}{\bf Super learners: richer set of learners (see Supplementary Tables \ref{table:ensemble.weights.disc.rich} and \ref{table:ensemble.weights.cont.rich})}\\
    Discrete-time SL: 10 periods, $L2$ loss  &0.200&18.8&72.4 &75.3\\
    Discrete-time SL: 10 periods, $loglik$ loss  &0.198&19.5&72.3&75.2\\
&&&&\\
    Continuous-time SL: Westling et al. 2023&0.196&20.5&72.2&75.6 \\
    \hline
    \end{tabular}
\end{table}

\section{Discussion}
\label{sec:discussion}

The SL is a powerful method for enabling flexible estimation of conditional expected outcome functions, which selects a best single model from a set of candidate learners or derives an optimal combination or `ensemble'. The candidate learners can include parametric and machine learning models, and the ensemble SL performs asymptotically at least as well as the single best performing model \citep{vdl_super_2007,polleyrose_superlearner_2011}. In many medical contexts the interest lies in a time-to-event outcome, and specifically in estimating or predicting a survival probability or a risk conditional on a set of individual characteristics, with estimation challenged by censoring. SLs for time-to-event outcomes have been developed, with new implementations appearing in recent years. However, the literature on the SL for time-to-event outcomes is technical, and somewhat disjointed, and a reader may find it challenging to gather together the full details of how these methods work and can be implemented based on the current literature. In this tutorial paper we have provided a general overview of the SL for time-to-event outcomes, followed by details of three specific implementations covering discrete-time and continuous-time versions of the SL. We have given a summary of how the methods can be implemented in R and illustrated the methods using an open access data set. R code is provided at \url{https://github.com/ruthkeogh/superlearner_survival_tutorial}, enabling others to reproduce our results. This includes by-hand code implementations that may help to provide additional clarification as well as helping users to adapt the methods in ways that may not be covered by the existing packages. 

\citet{polley_superlearner_2011} were the first to describe a SL for time-to-event outcomes based on a discrete-time approach. This approach has the advantage that it can make use of a wide range of candidate learners that exist for binary outcomes. While discretising time for the analysis could be considered undesirable, provided the discretisation is not too crude it is unlikely to have an important impact on the estimates. The discrete-time SL can use a loss function that does not involve the censoring distribution, or can use a loss function that requires a model for the censoring distribution to be pre-specified. The more recently proposed continuous-time SLs of \citet{westling_inference_2023} and \citet{munch_jossl_2025} have the advantage that they do not require a discretisation of time, and in our application we found these methods to give better predictive performance, though this is only one example. A drawback of the continuous-time approach is that there exist fewer candidate learners for continuous time-to-event data. The major advantage of the continuous-time approaches is that they allow for the censoring distribution (which is required for evaluating predictive performance and deriving the best fitting model or the ensemble via the loss function) to be estimated flexibly using a SL as part of the process. By contrast the discrete-time approach using the IPCW loss function requires the censoring distribution to be pre-specified, with this often being assumed not to depend on covariates, though it can also be extended to be conditional on covariates, for example using a separate SL.  

 \citet{munch_jossl_2025} noted that a drawback of their joint survival SL approach is that the loss function targets the conditional state occupation probabilities rather than the conditional survival probability, which is the target estimand. However their simulation studies indicated that good performance for the former translates into good performance for the latter. The loss functions used by \citet{westling_inference_2023} directly target the conditional survival probability, however \citet{munch_jossl_2025} pointed out that no general theoretical guarantees appear to have been given for the procedure of \citet{westling_inference_2023}. \citet{munch_jossl_2025} compared their approach with that of \citet{westling_inference_2023} in a simulation study, finding that the two methods performed similarly, with their joint survival SL approach often having slightly better performance. An important advantage of the approach of \citet{munch_jossl_2025} is that it extends to handle competing events, with details of this provided in their paper.

An alternative continuous-time SL was proposed by \citet{golmakani_superlearner_2020}. However, their approach is restricted to candidate learners that involve a linear predictor and have a proportional hazards form. It therefore excludes methods such as random survival forests. Their ensemble step is also somewhat more complicated than those explained for the methods summarised in this tutorial. For these reasons we decided to exclude this approach from our overview. Their methods can be implemented using code available at \url{https://github.com/kgolmakani/SuperLearner-Survival.git} and \url{https://github.com/kgolmakani/survSL.git}. There is another R package implementing a SL called \texttt{survivalSL} \citep{foucher_package_2025}, however we did not find detailed documentation at the time of writing. In terms of implementation, the \texttt{SuperLearner} package \citep{polley_package_2024} is currently the most well developed and flexible. 

In this paper we have focused on SL methods that are based on estimation of a discrete-time hazard, a survival function or a state occupation probability. An alternative approach is to target pseudo observations \citep{andersen_generalised_2003}, which enable censored survival data to be analysed as continuous outcomes. \citet{Sachs_ensemble_2019} described a SL using pseudo observations, which enables use of models for continuous outcomes in the set of candidate learners. They focused on a setting with competing risks and proposed a loss function based on an AUC. After calculating the pseudo observations they used the \texttt{SuperLearner} package with the pseudo observations as a continuous outcomes, using a bespoke loss function. Details of the implementation are provided at \url{https://github.com/sachsmc/pseupersims}. An application of this approach to our Rotterdam data example is summarised in Supplementary Section \ref{suppsec:pseudoobs}. \citet{cwiling_pseudo_2025} also considered a SL based on pseudo observations, focusing on estimation of restricted mean survival times using a quadratic loss function. A drawback of using pseudo observations in prediction is that it can give risk estimates that are negative or greater than 1. The prediction model also has to be estimated separately for different time horizons $\tau$, which can lead to survival probabilities that not monotonically decreasing over time.

As noted earlier, the SL is is type of `stacking algorithm' \citep{wolpert_stacked_1992, breiman_stacked_1996}. Use of stacking in the context of estimating a conditional survival function was proposed by \citet{wey_combining_2015}. \citet{wolock_framework_2024} proposed a `general survival stacking' approach, which involves writing the cumulative hazard in terms of probabilities, and uses weights to handle censoring and left-truncation. They refer to this as a global survival stacking approach, and it is implemented in the R package \texttt{survML}. \citet{wolock_framework_2024} compared their approach with the SL, including the methods of \citet{polley_superlearner_2011} (described as a `local' rather than `global' stacking method) and \citet{westling_inference_2023}. The method of \citet{wolock_framework_2024} handles left truncation as well as right censoring. The discrete-time method of \citet{polley_superlearner_2011} could handle left truncation easily, by making use only of time periods in which an individual is under observation, which was pointed out by \citet{wolock_framework_2024}. They also point out that the method of \citet{westling_inference_2023} does not handle left-truncation. It is also unclear how the SL of \citet{munch_jossl_2025} could adapt to accommodate left truncation. A unique feature of the SL of \citet{munch_jossl_2025} is that it is designed to handle competing events, which are often faced in time-to-event analyses. 

In our implementations of the methods our aim was not to conduct a formal comparison, but instead to illustrate how they can be applied. Some simulation based comparisons of the methods discussed in this paper have been conducted elsewhere \citep{munch_jossl_2025, wolock_framework_2024}. We considered using both a relatively simple set of candidate learners and a richer set of learners. In practice, using a diverse set of learners is a good idea, as it results in lower correlation between the errors from different models, which is desirable \citep{brown_diversity_2005}. For example, \citet{wey_combining_2015} suggested that including lots of similar models (e.g different Cox models, with different ways of entering the covariates) is not advisable, as the predictions from these models will be highly correlated. An overview of machine learning algorithms for survival data is provided by \citet{MLSA2025}. In practice, the choice between different SL methods, and the choice of candidate learners to be included, is likely to be based on practical considerations, including computational. 

We have placed an emphasis on the use of the SL as a tool for developing a prediction model. However, as noted earlier, estimation of conditional survival and censoring distributions is also a component of estimation procedures used in causal inference. For example, suppose that the causal estimand of interest is a marginal survival probability under a particular treatment strategy. In targeted maximum likelihood estimation (TMLE) the conditional survival and censoring distributions are nuisance models that need to estimated as part of a series of analysis steps used to estimate such a causal estimand, and the SL is typically used to estimate the nuisance models \citep{moore_rcts_2011}. \citet{westling_inference_2023} also used their SL approach to estimate nuisance models used as part of another type of doubly-robust estimation procedure for marginal survival curves under a particular treatment strategy. The methods described in this tutorial are therefore also relevant for those wishing to gain insights into the SL as used as a tool in causal inference estimation methods.

\vspace{1cm}

{\bf Funding}

RHK was funded by UK Research and Innovation (Future Leaders Fellowship MR/X015017/1). KDO was funded by a Royal Society-Welcome Trust Sir Henry Dale fellowship, grant number 218554/Z/19/Z. JMG received support from the Research Council of Norway (project number 352140).

\bibliographystyle{biorefs}
\bibliography{references}

\begin{thebibliography}{99}

\bibitem[Andersen \emph{and others}(2003)Andersen, Klein and Rosthøj]{andersen_generalised_2003}
\textsc{Andersen, P.K., Klein, J.P. and Rosthøj, S.} (2003).
\newblock Generalised linear models for correlated pseudo-observations, with applications to multi-state models.
\newblock {\em Biometrika\/}~\textbf{90}, 15--27.

\bibitem[Andersen \emph{and others}(2021)Andersen, Pohar~Perme, Houwelingen, Cook, Joly, Martinussen and Therneau]{andersen_analysis_2021}
\textsc{Andersen, P.K., Pohar~Perme, M., Houwelingen, H.C., Cook, R.J., Joly, P., Martinussen, T. and Therneau, T.M.} (2021).
\newblock Analysis of time‐to‐event for observational studies: Guidance to the use of intensity models.
\newblock {\em Statistics in Medicine\/}~\textbf{40}(1), 185--211.

\bibitem[Bender \emph{and others}(2018)Bender, Groll and Scheipl]{bender_generalized_2018}
\textsc{Bender, A., Groll, A. and Scheipl, F.} (2018).
\newblock A generalized additive model approach to time-to-event analysis.
\newblock {\em Statistical Modelling\/}~\textbf{18}(3-4), 299--321.

\bibitem[Breiman(1996)Breiman]{breiman_stacked_1996}
\textsc{Breiman, L.} (1996).
\newblock Stacked regressions.
\newblock {\em Mach Learn\/}~\textbf{24}, 49--64.

\bibitem[Brown \emph{and others}(2005)Brown, Wyatt, Harris and Yao]{brown_diversity_2005}
\textsc{Brown, G, Wyatt, J, Harris, R and Yao, X}. (2005).
\newblock Diversity creation methods: a survey and categorisation.
\newblock {\em Information Fusion\/}~\textbf{6}, 5--20.

\bibitem[Cox(1972)Cox]{cox_1972}
\textsc{Cox, D.R.} (1972).
\newblock Regression models and life-tables.
\newblock {\em Journal of the Royal Statistical Society. Series B (Methodological)\/}~\textbf{34}, 187--220.

\bibitem[Cwiling \emph{and others}(2025)Cwiling, Perduca and Bouaziz]{cwiling_pseudo_2025}
\textsc{Cwiling, A., Perduca, V. and Bouaziz, O.} (2025).
\newblock Pseudo-observations and super learner for the estimation of the restricted mean survival time.
\newblock {\em Lifetime Data Analysis\/}~\textbf{31}, 713--746.

\bibitem[Foucher and Sabathe(2025)Foucher and Sabathe]{foucher_package_2025}
\textsc{Foucher, Y. and Sabathe, C.} (2025).
\newblock {\em survivalSL: Super Learner for Survival Prediction from Censored Data\/}.

\bibitem[Golmakani and Polley(2020)Golmakani and Polley]{golmakani_superlearner_2020}
\textsc{Golmakani, M. and Polley, G.C.} (2020).
\newblock Super learner for survival data prediction.
\newblock {\em The International Journal of Biostatistics\/}~\textbf{16}(2).

\bibitem[Graf \emph{and others}(1999)Graf, Schmoor, Sauerbrei and Schumacher]{graf_assessment_1999}
\textsc{Graf, E., Schmoor, C., Sauerbrei, W. and Schumacher, M.} (1999).
\newblock Assessment and comparison of prognostic classification schemes for survival data.
\newblock {\em Statistics in Medicine\/}~\textbf{18}(17-18), 2529--2545.

\bibitem[Ishwaran \emph{and others}(2008)Ishwaran, Kogalur, Blackstone and Lauer]{ishwaran_random_2008}
\textsc{Ishwaran, H., Kogalur, U.B., Blackstone, E.H. and Lauer, M.S.} (2008).
\newblock Random survival forests.
\newblock {\em Ann. Appl. Stat.\/}~\textbf{2}(3), 841--860.

\bibitem[McLernon \emph{and others}(2022)McLernon, Giardiello, Calster, Wynants, van Geloven, van Smeden, Therneau and Steyerberg]{mclernon_assessing_2022}
\textsc{McLernon, D.J., Giardiello, D., Calster, B.~Van, Wynants, L., van Geloven, N., van Smeden, M., Therneau, T. and Steyerberg, E.W}. (2022).
\newblock Assessing performance and clinical usefulness in prediction models with survival outcomes: practical guidance for {Cox} proportional hazards models.
\newblock {\em Annals of Internal Medicine\/}, M22--0844.

\bibitem[Moore and van~der Laan(2011)Moore and van~der Laan]{moore_rcts_2011}
\textsc{Moore, K.~L. and van~der Laan, M.J.} (2011).
\newblock Rcts with time-to-event outcomes.
\newblock In: van~der Laan, M.J. and Rose, S. (editors), {\em Targeted Learning: Causal Inference for Observational and Experimental Data\/}, 1st edition, Chapter~17. New York: Springer, pp.\  259--269.

\bibitem[Munch and Gerds(2024)Munch and Gerds]{munch_statelearner_2024}
\textsc{Munch, A. and Gerds, T.A.} (2024).
\newblock The state learner -- a super learner for right-censored data.
\newblock {\em arXiv\/}, arXiv:2405.17259.

\bibitem[Munch and Gerds(2025)Munch and Gerds]{munch_jossl_2025}
\textsc{Munch, A. and Gerds, T.A.} (2025).
\newblock The joint survival super learner: A super learner for right-censored data.
\newblock {\em arXiv\/}, arXiv:2405.17259v2.

\bibitem[Naimi and Balzer(2018)Naimi and Balzer]{naimi_stacked_2018}
\textsc{Naimi, A.I. and Balzer, L.B.} (2018).
\newblock Stacked generalization: an introduction to super learning.
\newblock {\em European Journal of Epidemiology\/}~\textbf{33}(5), 459--464.

\bibitem[Polley \emph{and others}(2024)Polley, LeDell, Chris~Kennedy, Sam~Lendle and van~der Laan]{polley_package_2024}
\textsc{Polley, E., LeDell, E., Chris~Kennedy, C., Sam~Lendle, S. and van~der Laan, M.} (2024).
\newblock {\em SuperLearner: Super Learner Prediction\/}.
\newblock R package version 2.0-29.

\bibitem[Polley and van~der Laan(2011)Polley and van~der Laan]{polley_superlearner_2011}
\textsc{Polley, E.C. and van~der Laan, M.J.} (2011).
\newblock Super learning for right-censored data.
\newblock In: van~der Laan, M.J. and Rose, S. (editors), {\em Targeted Learning: Causal Inference for Observational and Experimental Data\/}, 1st edition, Chapter~16. New York: Springer, pp.\  249--258.

\bibitem[Polley \emph{and others}(2011)Polley, Rose and van~der Laan]{polleyrose_superlearner_2011}
\textsc{Polley, E.~C., Rose, S. and van~der Laan, M.~J.} (2011).
\newblock Super learning.
\newblock In: van~der Laan, M.~J. and Rose, S. (editors), {\em Targeted Learning: Causal Inference for Observational and Experimental Data\/}, 1st edition, Chapter~3. New York: Springer, pp.\  43--66.

\bibitem[Royston and Altman(2013)Royston and Altman]{royston_external_2013}
\textsc{Royston, P. and Altman, D.} (2013).
\newblock External validation of a cox prognostic model: principles and methods.
\newblock {\em BMC Medical Research Methodology\/}~\textbf{13}, 33.

\bibitem[Sachs \emph{and others}(2019)Sachs, Discacciati, Everhov, Olén and Gabriel]{Sachs_ensemble_2019}
\textsc{Sachs, M.C., Discacciati, A., Everhov, A.H, Olén, O. and Gabriel, E.E.} (2019).
\newblock Ensemble prediction of time-to-event outcomes with competing risks: A case-study of surgical complications in crohn's disease.
\newblock {\em Royal Statistical Society Series C: Applied Statistics\/}~\textbf{68}(5), 1431--1446.

\bibitem[Sonabend and Bender(2025)Sonabend and Bender]{MLSA2025}
\textsc{Sonabend, R. and Bender, A.} (2025).
\newblock {\em Machine Learning in Survival Analysis. https://www.mlsabook.com\/}.

\bibitem[Sonabend \emph{and others}(2024)Sonabend, Zobolas, Koipper, Burk and Bender]{sonabend_examining_2024}
\textsc{Sonabend, R., Zobolas, J., Koipper, P., Burk, L. and Bender, A.} (2024).
\newblock Examining properness in the external validation of survival models with squared and logarithmic losses.
\newblock {\em arXiv\/}, arXiv:2212.05260.

\bibitem[Tanner \emph{and others}(2020)Tanner, Sharples, Daniel and Keogh]{tanner_dynamic_2020}
\textsc{Tanner, K.T, Sharples, L.D, Daniel, R.M. and Keogh, R.H.} (2020).
\newblock Dynamic survival prediction combining landmarking with a machine learning ensemble: Methodology and empirical comparison.
\newblock {\em Journal of the Royal Statistical Society, Series A\/}~\textbf{184}(1), 3--30.

\bibitem[Therneau \emph{and others}(2024)Therneau, Lumley, Atkinson and Crowson]{therneau_survival_2024}
\textsc{Therneau, T.M., Lumley, T., Atkinson, E. and Crowson, C.} (2024).
\newblock {\em survival: Survival Analysis\/}.
\newblock R package version 3.7-0.

\bibitem[Tibshirani(1997)Tibshirani]{tibshirani_lasso_1997}
\textsc{Tibshirani, R.} (1997).
\newblock The lasso method for variable selection in the cox model.
\newblock {\em Statistics in Medicine\/}~\textbf{16}(4), 385--395.

\bibitem[Uno \emph{and others}(2011)Uno, Cai, Pencina, D’Agostino and Wei]{uno_c-statistics_2011}
\textsc{Uno, H., Cai, T., Pencina, M.J., D’Agostino, R.B. and Wei, L.~J.} (2011).
\newblock On the {C}-statistics for {Evaluating} {Overall} {Adequacy} of {Risk} {Prediction} {Procedures} with {Censored} {Survival} {Data}.
\newblock {\em Statistics in medicine\/}~\textbf{30}(10), 1105--1117.

\bibitem[van~der Laan \emph{and others}(2007)van~der Laan, Polley and Hubbard]{vdl_super_2007}
\textsc{van~der Laan, M.J., Polley, E.C. and Hubbard, A.E.} (2007).
\newblock Super learner.
\newblock {\em Stat Appl Genet Mol Biol\/}~\textbf{6}, Article 25.

\bibitem[Westling \emph{and others}(2023)Westling, Luedtke, Gilbert and Carone]{westling_inference_2023}
\textsc{Westling, T., Luedtke, A., Gilbert, P.~B. and Carone, M.} (2023).
\newblock Inference for treatment-specific survival curves using machine learning.
\newblock {\em Journal of the American Statistical Association\/}~\textbf{119}(546), 1541--1553.

\bibitem[Wey \emph{and others}(2015)Wey, Connett and Rudser]{wey_combining_2015}
\textsc{Wey, A., Connett, J. and Rudser, K.} (2015).
\newblock Combining parametric, semi-parametric, and non-parametric survival models with stacked survival models.
\newblock {\em Biostatistics\/}~\textbf{16}(3), 537--549.

\bibitem[Wolock \emph{and others}(2024)Wolock, Gilbert, Simon and Carone]{wolock_framework_2024}
\textsc{Wolock, C.~J., Gilbert, P.~B., Simon, N. and Carone, M.} (2024).
\newblock A framework for leveraging machine learning tools to estimate personalized survival curves.
\newblock {\em Journal of Computational and Graphical Statistics\/}~\textbf{33}(3), 1--11.

\bibitem[Wolpert(1992)Wolpert]{wolpert_stacked_1992}
\textsc{Wolpert, D.H.} (1992).
\newblock Stacked generalization.
\newblock {\em Neural networks\/}~\textbf{24}, 241--259.

\end{thebibliography}

\clearpage

\setcounter{section}{0}
\renewcommand{\thesection}{S\arabic{section}}

\setcounter{table}{0}
\renewcommand{\thetable}{S\arabic{table}}

\setcounter{figure}{0}

\renewcommand{\thefigure}{S\arabic{figure}}

\renewcommand{\theequation}{S\arabic{equation}}

\begin{center}
    \huge{{\bf Supplementary Materials}}
\end{center}

\begin{table}[!ht]
    \caption{Coefficients (or weights) for the candidate learners (based on a simple set of candidate learners) for discrete-time SL, for 10 or 100 time periods and using the $L2$ or $loglik$ loss function.}
    \label{table:ensemble.weights.disc.simple}
    \centering
    \begin{tabular}{lllll}
    \hline
       Candidate learner  &  10 periods &  10 periods&  100 periods &  100 periods\\
        & $L2$ loss &  $loglik$ loss&  $L2$ loss &  $loglik$ loss\\
        \hline
       Mean  &0.0145 &0.0166&0.2906&0.1668\\
       GLM &0&0.0402&0.6390&0.5901\\
       GLM - Lasso & 0.5469&0.5107&0&0\\
       Random forest & 0.4386&0.4325&0.0703&0.2430\\
           \hline
    \end{tabular}
\end{table}

\begin{table}[!ht]
    \caption{Coefficients (or weights) for the candidate learners (based on a simple set of candidate learners) for continuous-time SL of \cite{westling_inference_2023}, which fits SLs for both event and censoring distributions.}
    \label{table:ensemble.weights.cont.simple}
    \centering
    \begin{tabular}{lll}
    \hline
       Candidate learner  &  Coefficient: Event & Coefficient: Censoring \\
        \hline
       Kaplan-Meier  &0 &0\\
       Cox &0 &0\\
       Cox - Lasso & 0.184 &0.603\\
       Random survival forest & 0.816 &0.397\\
           \hline
    \end{tabular}
\end{table}

\begin{table}[]
    \caption{Coefficients (or weights) for the candidate learners (based on a richer set of candidate learners) for discrete-time SL, using 10 time periods and using the $L2$ loss function.}
    \label{table:ensemble.weights.disc.rich}
    \centering
    \begin{tabular}{ll}
    \hline
    Candidate learner & Coefficient \\
    \hline
      Mean   & 0.052 \\
      GLM   & 0\\
      GLM, Bayesian & 0\\
      GLM, Lasso ($\alpha=1$) & 0.029\\
      GLM, Ridge ($\alpha=0$)& 0\\
      GLM, Elastic net ($\alpha=0.5$)& 0\\
      GAM, 2 degrees of freedom& 0.166\\
      GAM, 3 degrees of freedom& 0.393\\
      GAM, 4 degrees of freedom& 0\\
      GAM, 5 degrees of freedom & 0\\
      XGBooost, max\_depth=3,shrinkage=0.1,minobspernode=10 & 0.149\\
      XGBooost, max\_depth=4,shrinkage=0.1,minobspernode=10 & 0\\
      XGBooost, max\_depth=4,shrinkage=0.1,minobspernode=1 & 0\\
      XGBooost, max\_depth=6,shrinkage=0.1,minobspernode=10 & 0\\
      XGBooost, max\_depth=6,shrinkage=0.1,minobspernode=1 & 0\\
      XGBooost, max\_depth=6,shrinkage=0.3,minobspernode=10 & 0\\
      XGBooost, max\_depth=6,shrinkage=0.3,minobspernode=1 & 0\\
      Random forest (using ranger), minnodesize=1 & 0\\
      Random forest (using ranger), minnodesize=2 &0\\
      Random forest (using ranger), minnodesize=5 &0.212\\
      \hline
    \end{tabular}
\end{table}

\begin{table}[!ht]
    \caption{Coefficients (or weights) for the candidate learners (based on a richer set of candidate learners) for the continuous-time SL of \cite{westling_inference_2023}, which fits SLs for both event and censoring distributions.}
    \label{table:ensemble.weights.cont.rich}
    \centering
    \begin{tabular}{lll}
    \hline
       Candidate learner  &  Coefficient: Event &Coefficient: Censoring\\
        \hline
       Kaplan-Meier  &0 & 0 \\
       Cox &0 &0\\
       Cox - Lasso & 0 &0.477\\
       Cox, with GAMs &0.592& 0\\
       Random survival forest & 0.408 &0.246\\
       Exponential model & 0 &0\\
       Weibull model &0 &0\\
       Log-logistic model & 0 &0.025\\
       Piecewise constant hazard model &0 &0.252\\     
           \hline
    \end{tabular}
\end{table}

\begin{table}[!ht]
    \caption{Results from the application of the discrete-time SL (using 100 time periods) to the Rotterdam data, using a simple set of candidate learners SLs. These results were obtained using the by-hand implementations of the methods.}
    \label{table:prediction.measures.byhand}
    \centering
    \begin{tabular}{lllll}
    \hline
    Prediction method&Brier score&Scaled Brier&C-index&AUCt\\
    \hline
    Discrete-time SL: 100 periods, $L2$ loss &0.213 &13.5  &72.1 &75.3 \\
    Discrete-time SL: 100 periods, $loglik$ loss &0.201 &18.3  &71.9 &74.8 \\
    Discrete-time SL: 100 periods, IPCW loss &0.224 &9.1  &70.9 &73.7 \\
    \hline
    \end{tabular}
\end{table}

\clearpage

\section{SL based on pseudo observations}
\label{suppsec:pseudoobs}

We applied the SL approach of \citet{Sachs_ensemble_2019} based on pseudo observations to the Rotterdam data illustration. To focus on our simpler setting with no competing risks we made minor modifications to the functions provided at \url{https://github.com/sachsmc/pseupersims}, which are required to define the loss function. In the estimation we made use of pseudo observations estimated at times $t=1,2,\ldots,10$ years, which were combined in a stacked data set. We used the same set of candidate learners as used in the code of \citet{Sachs_ensemble_2019}, which include a GLM, GAM, random forest, support vector machine, adaptive polynomial spline regression, recursive partitioning and regression trees, and XGBoost with different tuning parameters. Predictive performance of the SL was evaluated in the test data. The Brier score was 0.204, which corresponded to a scaled Brier score of 17.3. The C-index and AUC in the test data were respectively 70.8 and 75.7. These results are similar to the estimates found for other SLs as shown in Table \ref{table:prediction.measures}. A drawback of using pseudo observations in prediction is that it can give risk estimates that are negative or greater than 1. This happened in our application and we truncated estimates to 0 or 1. 

The code used to implement the pseudo observations SL approach is available at \url{https://github.com/ruthkeogh/superlearner_survival_tutorial}. 

\end{document}